\documentclass[preprint,unsortedaddress,superscriptaddress]{revtex4}
\usepackage{amsmath,color}
\usepackage[pdftex]{graphicx}
\allowdisplaybreaks

\begin{document}
\title{Reverse Doppler Effect of Sound}

\vspace{\baselineskip}

\author{Sam Hyeon Lee}
\affiliation{Institute of Physics and Applied Physics, Yonsei
University, Seoul 120-749, Korea}
\author{Choon Mahn Park}
\affiliation{AEE center, Anyang University, Anyang, 430-714, Korea}
\author{Yong Mun Seo}
\affiliation{Department of Physics, Myongji University, Yongin
449-728, Korea}
\author{Zhi Guo Wang}
\affiliation{Department of Physics, Tongji University, Shanghai
200092, People's Republic of China}

\author{Chul Koo Kim\footnote{To whom correspondence should be addressed. E-mail:
ckkim@yonsei.ac.kr}}
\affiliation{Institute of Physics and Applied
Physics, Yonsei University, Seoul 120-749, Korea}

\date\today

\vspace{\baselineskip}

\begin{abstract}

We report observation of reverse Doppler effect in a double negative
acoustic metamaterial. The metamaterial exhibited negative phase
velocity and positive group velocity. The dispersion relation is
such that the wavelength corresponding to higher frequency is
longer. We observed that the frequency was down-shifted for the
approaching source, and up-shifted when the source receded.

\end{abstract}

\maketitle

Metamaterials with negative constitutive parameters~\cite{1,2,3,4,5}
brought many phenomena previously regarded impossible into
reality~\cite{6,8,9,10}. Electromagnetic waves have been
demonstrated to propagate with negative phase velocity in the double
negative (DNG) matematerials which have negative electric
permittivity and negative magnetic permeability simultaneously. For
acoustic waves, negative phase velocity is expected if density and
modulus can be made simultaneously negative~\cite{11}. We combined
negative density material~\cite{12,13} and negative modulus
material~\cite{14,15} to fabricate an acoustic metamaterial that
exhibits the acoustic DNG property. The negative phase velocity
introduces a dispersion relation in which the wavelengths
corresponding to higher frequencies become longer. By measuring the
frequency shifts in the DNG metamaterial, we observed the reverse
Doppler effect of sound, where the frequency was down-shifted for
the approaching source and up-shifted when the source receded.

In comparison with the numerous theoretical and experimental
research results on electromagnetic metamaterials, research efforts
on the acoustic metamaterials are relatively few and mainly
concentrated on theoretical aspects~\cite{11,17,18,19,20,21,22}.
Only, recently, Fang \emph{et al.} reported realization of negative
modulus in a metamaterial consisting of Helmholtz
resonators~\cite{12}. Also negative mass has been realized using a
membrane by Yang \emph{et al.}~\cite{14}. The authors also have
investigated single negative acoustic metamaterials~\cite{13,15}.
Here, we present fabrication of a new acoustic DNG metamaterial by
combining the two single negative metamaterials. Fig. 1a shows
schematically the structure of a negative density
material~\cite{13}, which has a regular array of tensioned thin
membranes placed inside a tube. The metamaterial shown in Fig. 1b
exhibits negative modulus due to the motion of air column in the
side holes~\cite{15}. Fig. 1c shows the acoustic DNG metamaterial
with the membranes and side holes alternately placed along the tube.

\begin{figure}
\begin{center}
\includegraphics*[width=1.0\columnwidth]{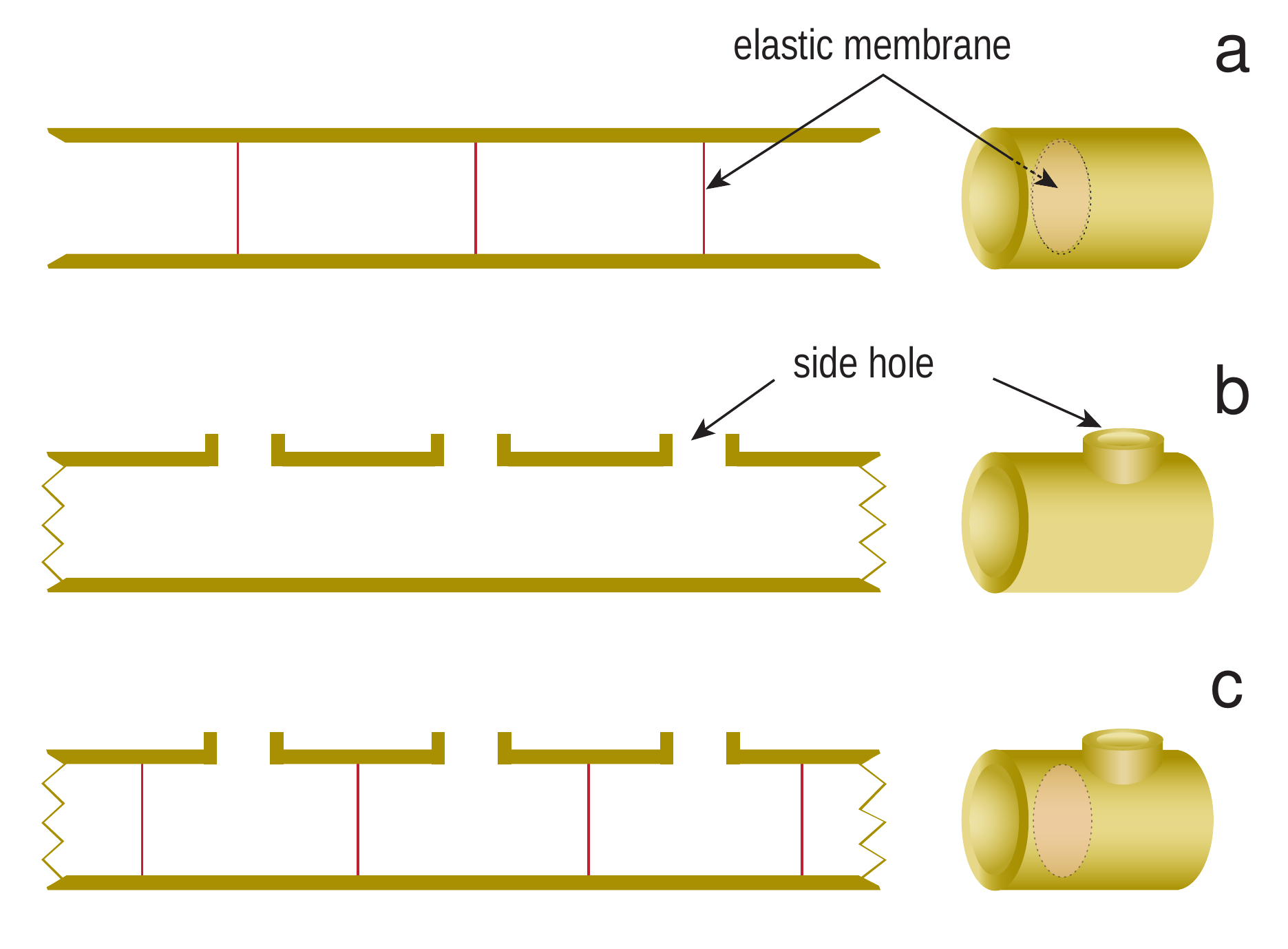}
\end{center}
\caption{Structures of metamaterials (\textbf{a} - \textbf{c}).
\textbf{a}, One-dimensional structure consisting of thin tensioned
elastic membranes in a tube. Negative effective density is observed
in this system. \textbf{b}, A tube with an array of side holes that
exhibits negative effective modulus. \textbf{c}, An acoustic DNG
structure with both membranes and side holes.} \label{fig:diag}
\end{figure}

Acoustic double negativity of this structure is due to the actions
of the membranes and side holes. The membrane forms a paraboloid
when pushed by the fluid, and when the spacing between the membranes
is much smaller than the wavelength (the long-wavelength limit), the
restoring force from the tension gives the static pressure gradient
${\bigtriangledown}p = -\kappa  \vec{\xi}$ where $p$, $\kappa$, and
$\vec{\xi}$  are the pressure, the elastic modulus, and the
displacement of the fluid, respectively. In the dynamic case, due to
the force from the membranes, Newton's equation becomes $-
\bigtriangledown p = {\rho}'
\partial \vec{u}/\partial t + \kappa \vec{\xi}$, where $\vec{u} =
\partial \vec{\xi}/\partial t$ is the longitudinal velocity of the
fluid. The volume-averaged density ${\rho}'$ of the fluid (air in
the present case) and the membranes, giving the inertia term in the
equation, is significantly different from the dynamic effective-mass
density ${\rho}_{eff}$ defined below. Using the harmonic expressions
$\vec{u}(x,t) = \vec{U} e^{-i\omega t}$, the equation can be written
in a convenient form,
\begin{equation} \label{eq:sound-2}
-{\bigtriangledown}p = \left( {\rho}'-\frac{\kappa}{\omega
^2}\right) \frac{\partial \vec{u}}{\partial t} .
\end{equation}
The proportionality constant of the acceleration to the pressure
gradient force is defined as the effective density,
\begin{equation} \label{eq:sound-3}
\rho_{eff} = {\rho}' - \frac{\kappa}{\omega^2} = {\rho}' \left(
1-\frac{\omega^2_{MEM}}{\omega ^2}\right),
\end{equation}
which becomes negative below the critical frequency $\omega_{MEM} =
\sqrt{\kappa/{\rho}'}$  $\left( f = \omega/2 \pi \right)$.

The longitudinal wave motion in the tube is additionally affected by
the motion of air moving in and out through the side holes (SH). The
air column in each hole has a mass given by $M = \rho_0 l'
S$~\cite{23}, where $l'$ is the effective length and $S$ is the
cross sectional area of the hole. The mass of air moves in and out
with velocity $v$, driven by the pressure $p$ in the tube according
to Newton's law $p S = M dv/dt$. When there are $n$ holes per unit
length, we can define the SH-mass-density and the
SH-cross-sectional-density as $\rho_{SH} = n M$ and $\sigma_{SH} = n
S$, respectively. The SH acts as a sink that modifies the continuity
equation in the tube: $- (1/B)
\partial p/\partial t = \bigtriangledown \cdot \vec{u} + \left(
\sigma_{SH}/A \right) v$, where $A$ is the cross section of the
tube. Using the harmonic expressions, this can be simplified to
\begin{equation} \label{eq:sound-5}
{\bigtriangledown}\cdot \vec{u} = - \left(
\frac{1}{B}-\frac{\sigma^2_{SH}}{\rho_{SH} A \omega ^2}\right)
\frac{\partial p}{\partial t} .
\end{equation}
The proportionality constant of the expansion $\left(
{\bigtriangledown}\cdot \vec{u} \right)$ to the pressure drop
$\left( -{\partial p}/{\partial t} \right)$ is defined as the
effective modulus,
\begin{equation} \label{eq:sound-6}
B_{eff} = \left( \frac{1}{B}-\frac{\sigma^2_{SH}}{\rho_{SH} A \omega
^2}\right)^{-1} = B \left( 1-\frac{\omega^2_{SH}}{\omega
^2}\right)^{-1}
\end{equation}
where $\omega_{SH} = {\left( {B\sigma^2_{SH}}/{A \rho_{SH}}\right)
}^{1/2}$. Thus, the system is described by the dynamic and
continuity equations $-\bigtriangledown p = {\rho}_{eff}\left(
\partial \vec{u}/ \partial t \right)$ and $\bigtriangledown \cdot \vec{u} =
-\left({1}/{B_{eff}}\right)\left(
\partial p/ \partial t \right)$, with the effective density and modulus given
by equations \ref{eq:sound-3} and \ref{eq:sound-6}. The resulting
wave equation gives the phase velocity,
\begin{equation} \label{eq:sound-7}
v_{ph} = \pm \sqrt{\frac{B_{eff}}{\rho_{eff}}} = \pm
\sqrt{\frac{B}{{\rho}'\left(1-\omega^2_{MEM}/\omega^2
\right)\left(1-\omega^2_{SH}/\omega^2 \right)}}.
\end{equation}

For the present system the critical frequencies, $f_{SH}$ and
$f_{MEM}$ are found to be 440 and 765 Hz respectively. That is, the
effective modulus is negative below 440 Hz and the effective density
is negative below 765 Hz. Thus, the system is DNG below 440 Hz and
the phase velocity in equation \ref{eq:sound-7} takes the negative
sign. As a consequence, there are three frequency ranges: In the DNG
range below 440 Hz, the acoustic waves are expected to propagate
with negative phase velocities. In the $\rho-$negative ($\rho$-NG)
range from 440 to 765 Hz, the phase velocity becomes imaginary and
the waves do not propagate. In the double positive (DPS) range above
765 Hz, the waves propagate with positive velocities. These
predictions are experimentally verified as shown below.

\begin{figure}
\begin{center}
\includegraphics*[width=0.6\columnwidth]{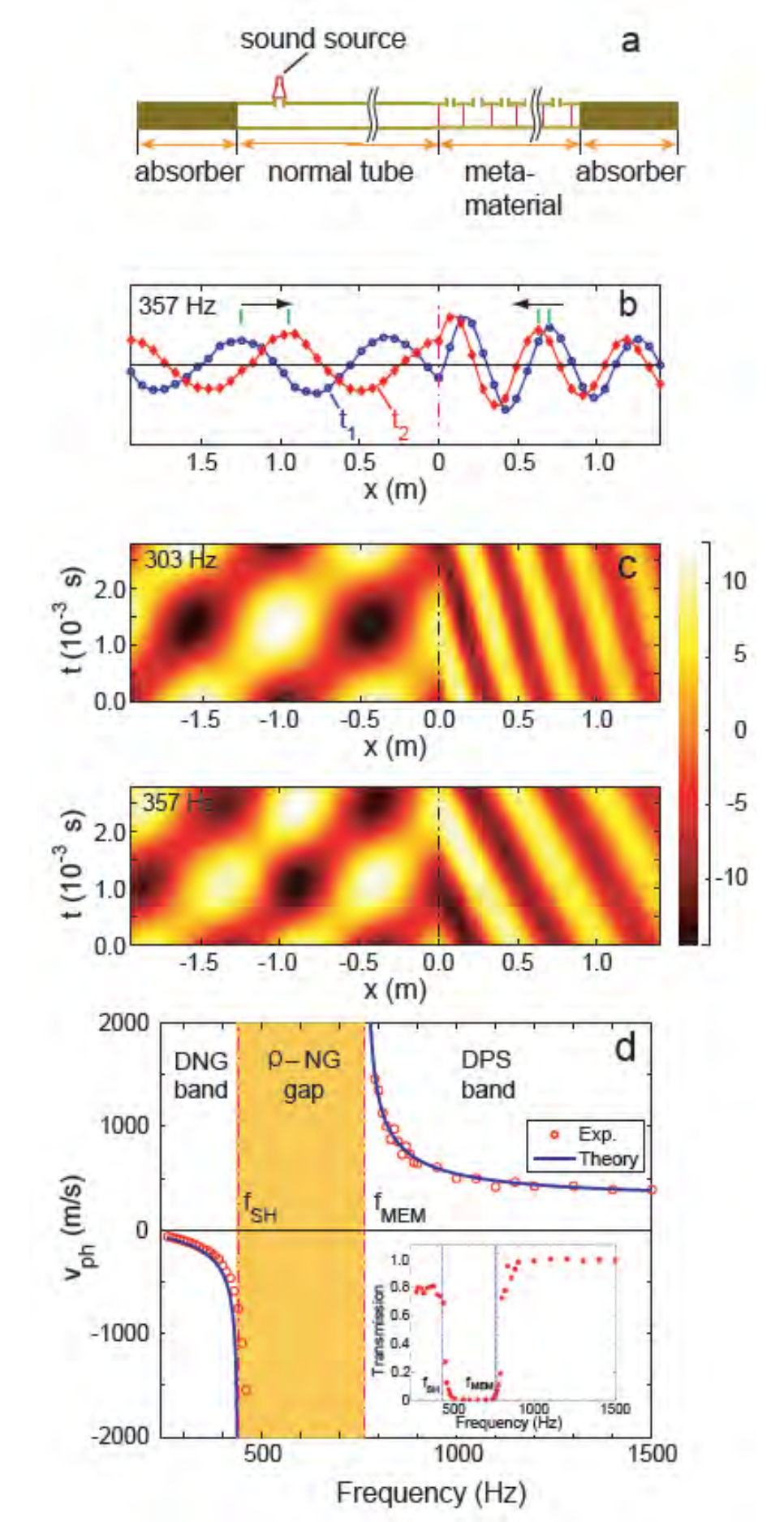}
\end{center}
\caption{Properties of wave propagations in the acoustic DNG
material. \textbf{a}, Experimental setup for the transmission and
phase velocity measurements. \textbf{b}, 'Snapshots' of measured
pressure distribution showing back-ward wave propagation in the
metamaterial $(x > 0)$. \textbf{c}, Characteristic diagrams of
pressure measurements for the frequencies 303 and 357 Hz. Negative
slopes of the wave-paths in the metamaterial sides $(x > 0)$
indicate negative phase velocities. \textbf{d}, Transmission (inset)
and phase velocities.} \label{fig:exp}
\end{figure}

Pressure was measured as a function of time and position on both the
normal tube side and the metamaterial side using the experimental
setup in Fig. 2a. Typical results for the frequency of 357 Hz are
shown in Fig. 2b. Two 'snapshots' of the measured pressure
distribution at times $t_1$ = 1.6 ms and $t_2$ = 2.0 ms are shown.
It can be seen that on the normal tube side, the wave proceeded
forward, but on the metamaterial side, the wave propagated backward,
as indicated by the arrows. Clearly, the wave in the metamaterial
propagated in a direction antiparallel to the energy flow. This
confirms the theoretical prediction of negative phase velocity.

The overall picture of the wave motion can be seen in the
characteristic diagram of $p(x, t)$  in which the axes are the
independent variables $t$ and $x$ (Fig. 2c). The position of the
boundary is indicated by broken lines at $x$ = 0 in the diagram. For
both of the exemplar frequencies (303 and 357 Hz), the slopes of the
paths followed by the wavecrests in the $(x, t)$ planes were
negative in the metamaterial $(x > 0)$. The phase velocities
determined from the slopes were approximately -110 m/s and -200 m/s
for 303 and 357 Hz, respectively. The wavelengths obtained from
these values were $\lambda$ = 0.36 m for 303 Hz and $\lambda$ = 0.56
m for 357 Hz. The wavelength for the higher frequency was longer
than that for the lower frequency. This special dispersion relation
is responsible for reverse Doppler effect discussed below. In the
normal tube side $(x < 0)$, the pattern of partial standing waves
caused by the reflected waves from the boundary was clearly seen.
The sound velocities in the normal tube obtained from the slope of
the patterns for the two frequencies had the same value, 340 m/s.

Comparison between the theory and experiment are shown in Fig. 2d.
Theoretically expected single negative gap is experimentally
confirmed by the transmission data (inset). In the DNG and DPS pass
bands, the phase velocities experimentally determined agree well
with the theoretical values. Therefore, it is clear that our
calculation gives an accurate description of the behavior of the
phase velocity in the frequency range from 250 to 1500 Hz.

One of the most prominent features of the DNG metamaterials is the
reversal of Doppler effect~\cite{1,5}. However, explicit observation
of reverse Doppler shift in metamaterials has not been reported,
because of difficulties of phase shift measurements inside of the
metamaterials. Thus, demonstration of reverse Doppler effect was so
far either carried out using a circuit configuration~\cite{24} or
reflecting a wave from a moving discontinuity~\cite{25}. Here we
report a direct observation of the reverse acoustic Doppler effect
in our DNG material. A sound source was made to move above the side
holes as shown in Fig. 3a, so that the sound propagated into the
tube through the holes. When the signal from the stationary detector
in the tube was sent to a loudspeaker, we heard a reversed Doppler
tone-shift. While the familiar Doppler tone-shift is from a high to
a low pitch as the sound source passes by the listener, the shift we
heard went from a low to a high pitch. The signal from the detector
for a typical measurement can be seen in Fig. 3b, where we moved the
source emitting 350 Hz sound at a speed of 5 m/s. The amplitude dip
in the middle is due to the reversed direction of the side hole at
the position of the detector (Fig. 3a). The numbers of oscillations
for the approaching and receding sound sources in a time interval of
0.05 s are compared in the expanded view (Fig. 3b). The frequencies
before and after the dip were determined to be 340 and 360 Hz,
respectively. Obviously, the frequency was down-shifted for the
approaching source, and up-shifted when the source receded. Observed
Doppler shift agrees well with the measured phase velocity.

\begin{figure}
\begin{center}
\includegraphics*[width=0.7\columnwidth]{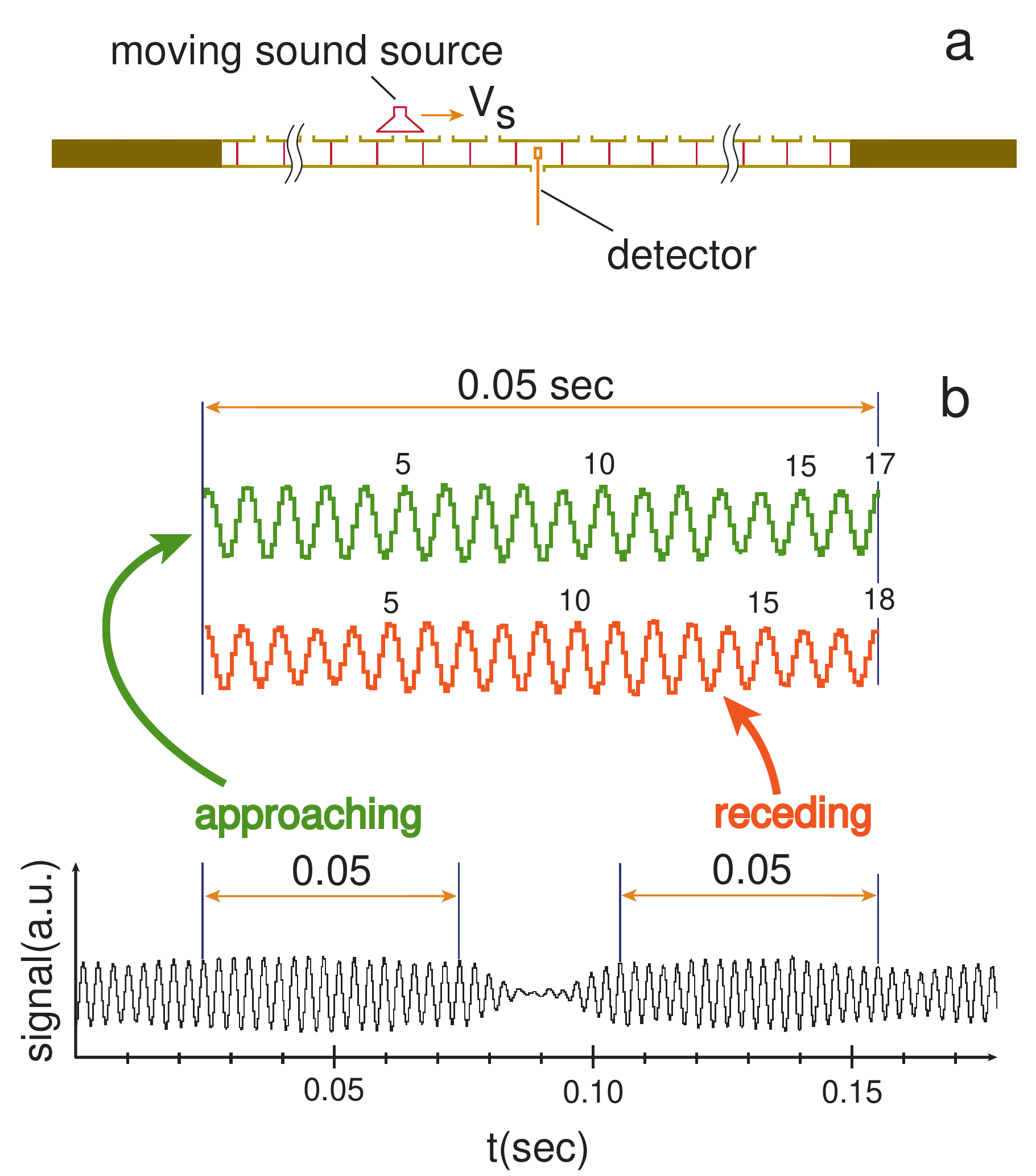}
\end{center}
\caption{ Observation of the reverse Doppler effect. \textbf{a},
Experimental setup for the Doppler experiment. The sound from the
moving source propagates into the tube through the side holes and is
detected by the stationary detector in the tube. \textbf{b},
Recorded signal from the stationery detector in the metamaterial. In
a time interval of 0.05 s, there were 17 periods of oscillation when
the source approached, while there were 18 when the source receded.}
\label{fig:doppler}
\end{figure}

One physically meaningful question is whether sound waves in front
of the moving source in our DNG material are compressed or expanded.
From Fig. 2d, in the DNG band, it can be seen that the acoustic
wavelength increases with frequency, i. e. high frequency
corresponds to longer wavelength. Therefore, from the fact that the
frequency was down-shifted, it can be inferred that the wavelength
in front of a moving source became shorter. So, we conclude that
even though the phase velocity is negative in the metamaterial, the
waves in front of the moving source is compressed, and that in the
rear is expanded.

In conclusion, we fabricated a DNG acoustic metamaterial consisting
of membranes and side holes on a tube. We observed  backward-wave
propagation and the reverse Doppler effect in this metamaterial.

\phantom{aaa} \phantom{aaa}

\textbf{METHODS} \\
\rule{\linewidth}{0.5pt}

The metamaterial shown in Fig. 1c has the inner diameter of about
32.3 mm with the length of the unit cell $d(= 1/n)$ = 70 mm. The
average density of the air loaded with the membrane in the tube was
${\rho}' \sim 1.34$ kg/m$^3$. The modulus from the tension of the
membranes was found to be $\kappa = 3.1 \times 10^7$ N/m$^4$. From
these values a critical frequency for $\omega_{MEM} =
\sqrt{\kappa/{\rho}'}$ was calculated to be about 765 Hz. The side
holes had a diameter of 10.0 mm ($S$ = 78.5 mm$^2$, $A$ = 821
mm$^2$). Using the value $M = 1.99 \times 10^{-6}$ kg, we obtained
the parameters, $\rho_{SH} = M/d = 2.84 \times 10^{-5}$ kg/m, and
$\sigma_{SH} = S/d = 1.12 \times 10^{-3}$ m. The critical frequency
for $\omega_{SH} = \sqrt{B \sigma_{SH}^2 /A \rho_{SH}}$, was
calculated to be about 440 Hz (where $B = 1.42 \times 10^5$ Pa).

The experimental setup in Fig. 2a consists of a normal tube on the
left and a 2 m long DNG metamaterial on the right. The absorbers at
both ends completely absorb the acoustic energy, preventing any
reflection and, thus, the system behaves as if it extends to
infinity. This eliminates concerns about the effect of the finite
number of cells used in the experiment, as well as the interference
effect from the reflected waves. The sound source injects acoustic
energy into the tube through a small hole, generating incident waves
propagating to the right. At the boundary, a portion of the incident
energy is reflected and the rest is transmitted into the
metamaterial region. On the metamaterial side, the transmitted
acoustic energy flows steadily to the right until it hits the
absorber. Waves propagating in this one-dimensional setup are
monitored by measuring pressures using miniature microphone.

Sound source in Fig. 2a was a miniature speaker (RP-HV102, Panasonic
Inc., 15 mm in diameter). Sound source in Fig. 3a, with a diameter
of 50 mm, was made to move 10 mm above the metamaterial tube with an
adjustable speed. The detectors were high sensitivity miniature
condenser-type microphones (MS-9600, Neosonic Inc., 7 mm in
diameter) with a thin connecting lead. The speakers were driven by
an arbitrary function generator (33220A, Agilent).

\phantom{aaa} \phantom{aaa}

\textbf{Acknowledgements}

The research was partially supported by The Korea Science and
Engineering Foundation (KOSEF R01-2006-000-10083-0).

\phantom{aaa} \phantom{aaa}

\textbf{Competing financial interests}

The authors declare that they have no competing financial interests.

\end{document}